# The turbulent formation of stars

Christoph Federrath

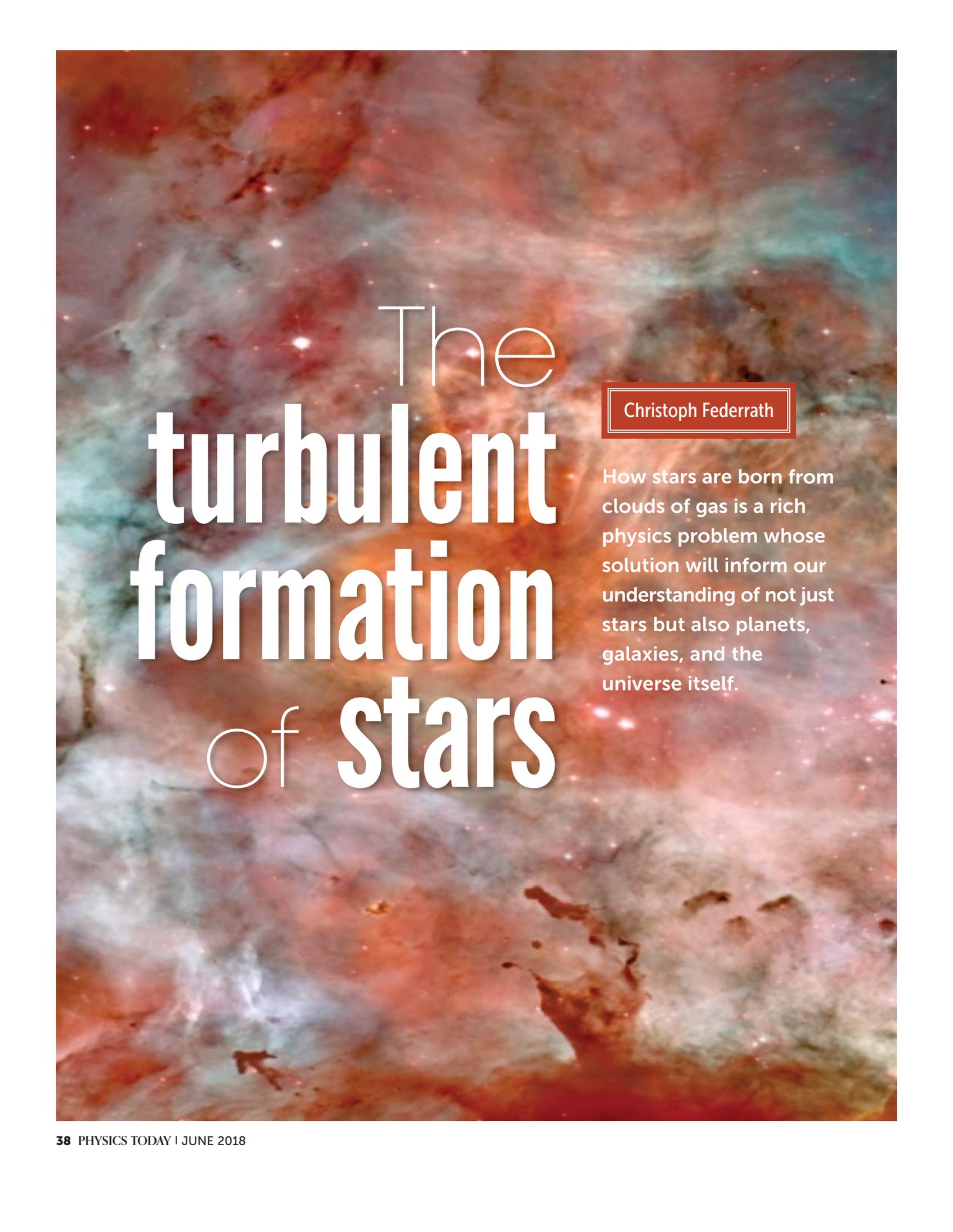

# The turbulent formation of stars

Christoph Federrath

**How stars are born from clouds of gas is a rich physics problem whose solution will inform our understanding of not just stars but also planets, galaxies, and the universe itself.**

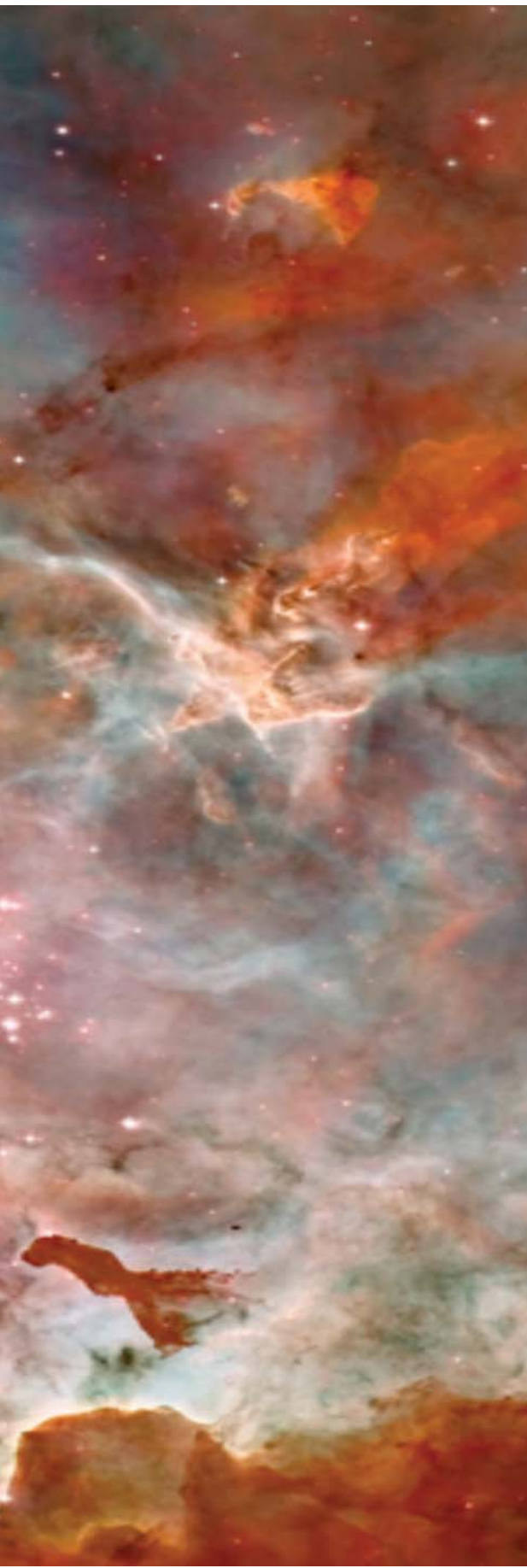

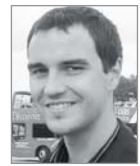

**Christoph Federrath** is a member of the faculty of the research school of astronomy and astrophysics at the Australian National University in Canberra.

Star formation is stupendously inefficient. Take the Milky Way. Our galaxy contains about a billion solar masses of fresh gas available to form stars—and yet it produces only one solar mass of new stars a year. Accounting for that inefficiency is one of the biggest challenges of modern astrophysics. Why should we care about star formation? Because the process powers the evolution of galaxies and sets the initial conditions for planet formation and thus, ultimately, for life.

Among star formation's most important physical ingredients is gravity. It can pull concentrations of gas together to create such high densities that nuclear fusion can start. But there is a big problem with gravity. When we look at typical gas clouds in the Milky Way, we find that they are already quite dense. If the clouds' densities and masses were all that mattered, star formation in our galaxy would be 10–100 times as fast as it actually is.

Over the past 30 years, researchers have been trying to understand why star formation is so slow. Basically, some energy source or a force must exist that counteracts what would otherwise be the clouds' fast gravitational collapse. But what is it? At first, magnetic fields were thought to provide the major balancing force against gravity. Clouds are indeed observed to have magnetic fields. However, the idea of magnetic field–regulated star formation faced several problems. Most important, if a magnetic field is strong enough to balance gravity initially, it will retain its predominance forever, and thus no stars will form. Theorists had to invoke finely tuned processes, such as ambipolar diffusion (the drift of ions and neutral atoms), to allow the clouds to lose magnetic flux on just the right time scales to explain the observed low star-formation rate. Then came the realization that magnetic fields, given their observed spatial distribution in the clouds, cannot retard star formation to the degree that the theory predicted.

At the same time, astronomers inferred that the internal motion of the gas in clouds is highly chaotic—that is, turbulent. The turbulence is evident from the large Doppler broadening of molecular lines observed throughout the clouds, notably by the most powerful radio telescope on Earth, the Atacama Large Millimeter/Submillimeter Array (ALMA). The kinetic energy that the turbulence carries is comparable to the gravitational energy of the clouds. That insight led to the main topic of this article: the recent and successful theory of turbulence-regulated star formation.[1]

The basic idea behind the theory is that turbulence plays a dual role. On one hand, by kicking the gas around, the turbulence makes it harder for gravity to collapse the clouds, which solves the original problem of accounting for the slow rate of star formation. On the other hand, the turbulence is supersonic—that is, the gas travels faster than pressure fluctuations

NASA, ESA, N. SMITH (UC BERKELEY), HUBBLE HERITAGE TEAM (STSCI/AURA), AND NOAO/AURA/NSF

can propagate. As a result, the gas experiences shocks and strong local compressions, which are necessary to seed gravitational collapse. The upshot: Turbulence kick-starts star formation that takes place in localized regions of the cloud that turbulence helped to create.

Once stars form, their winds blow material into the interstellar medium (ISM). More material is fed back into the ISM when massive stars die in supernova explosions. Crucially, the stellar feedback replenishes and sustains the ubiquitous turbulence and, with it, star formation itself. Turbulence helps initiate the formation of stars, which feeds material and energy back into the ISM and drives further turbulence. But large-scale dynamical processes, such as the shear induced by galactic rotation and the accretion of gas from outside the galaxy, also drive turbulence. In the turbulence-regulated picture of star formation, it is important to identify and understand the various drivers.

Turbulence, however, is not the whole story. Gravity, magnetic fields, and the nonturbulent manifestations of stellar feedback all play a role in star formation. Although the relative importance of those phenomena is not well understood, it seems likely that only their complex interplay accounts for the star-formation rates we observe, a view supported by the simulations shown in figure 1.

## The role of turbulence

Most stellar feedbacks, such as supernova explosions and stellar winds, and some galactic processes, such as accretion and spiral shocks, primarily drive compressive (curl-free) modes of turbulence. By contrast, shear and magnetorotational instability primarily drive solenoidal (divergence-free) modes. In reality, turbulence drivers excite a mixture of both compressible and solenoidal modes. For example, the jets that shoot from the poles of some young stars rotate when they propagate. As the jets drill through the ISM, the shear flows that develop between their inner and outer parts drive solenoidal modes. At the same time, the bow shock at the tip of the jet drives compressive modes.

Compressive modes are more effective at seeding star formation. In fact, they produce star-formation rates up to an order of magnitude higher than solenoidal modes would do for otherwise similar cloud parameters.[2] That difference helps to explain the relatively inefficient star formation at the center of our galaxy. There, despite high gas densities, strong shear flows induce solenoidal modes, which reduce the star-formation rate.[3] In particular, the theory of turbulence-regulated star formation predicts a rate of only 0.01 solar mass per year in the so-called Brick cloud near the galactic center, a rate that is indeed observed.[4] On larger, galaxy scales, shear can also contribute to reducing and regulating star formation.[5]

Figure 2 illustrates the crucial effect of turbulence mode on the star-formation rate. Two simulations of star-cluster formation are compared: one with solenoidal driving and the other with compressive driving. Everything else—the mass, the size, the strength of the turbulence, and so on—is the same in both simulations. The star-formation rate is a factor of 15–20 higher with compressive turbulence than with solenoidal turbulence.

Recent observations of clouds in the Milky Way[6] and numerical simulations of molecular cloud formation[7] indicate that what drives the turbulence in any given cloud could be purely compressive, purely solenoidal, or an arbitrary mix of the two. More observational studies are needed to characterize the turbulent modes in different clouds and in different galactic environments.

## How massive are new stars?

The initial mass of a star predestines its life, its death, and the evolution of material around it. The more massive a star, the more energy it puts out and the faster it will consume its nuclear fuel. The most massive stars radiate so powerfully that they ionize the surrounding gas. Collectively, massive stars emit most of the light we observe from distant galaxies. Because of the huge effect that massive stars have on galaxy evolution, we need to understand what processes determine a star's birth mass.

Astronomers call the observed mass distribution of newly formed stars the initial mass function (IMF). We know from observational surveys that most stars have masses less than about half that of the Sun. Stars with lower and higher masses are rarer. At the high-mass tail of the IMF—up to a few hundred solar masses—the number of stars of mass $M$ declines[8] in proportion to the power-law $M^{-1.35}$. Just how this seemingly universal power-law tail and the observed peak at about 0.1 solar masses arise is one of the most challenging problems in astrophysics.

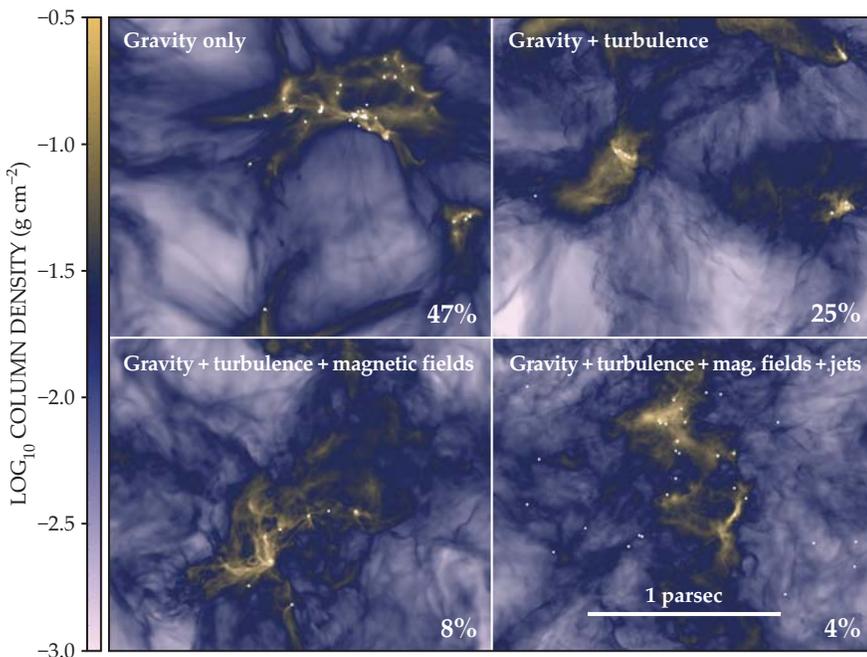

**FIGURE 1. EXPLAINING STAR FORMATION'S INEFFICIENCY** requires invoking physical processes beyond gravitational collapse. The four panels show numerical simulations of star formation in the same clouds, but with increasingly rich physics.[15] White dots denote stars. The percentages give the fraction of the cloud's original mass that formed stars within or normalized to the cloud's free-fall time—that is, the time that it would take the cloud to collapse under its own gravity. If the fraction is 100%, the whole cloud has turned its gas into stars within one free-fall time. A fraction of 1% would indicate that it would take 100 free-fall times to convert all the cloud's gas into stars. Only the combination of gravity, turbulence, magnetic fields, and stellar feedback from jets yields rates that match observations of real clouds.

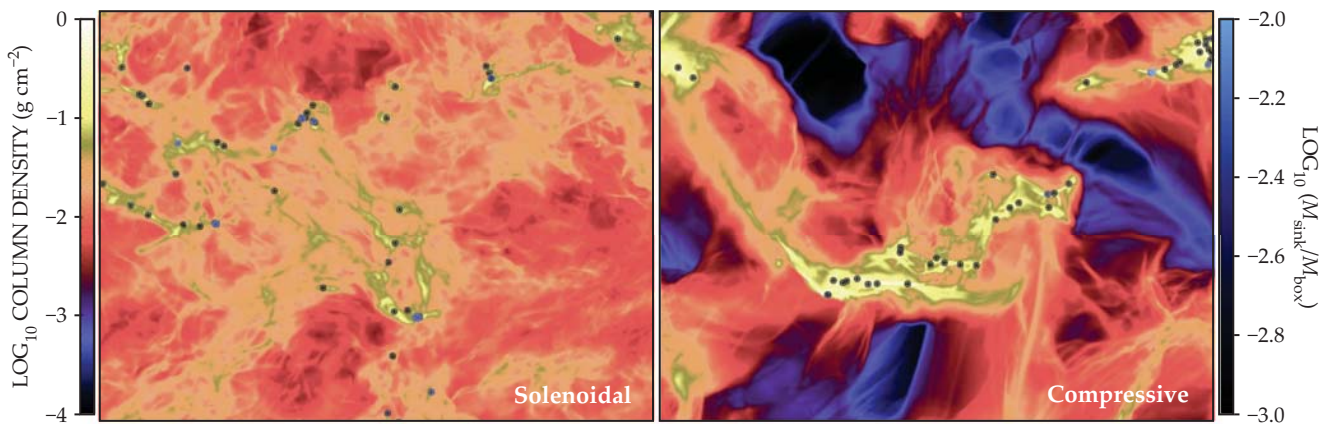

**FIGURE 2. WHETHER THE TURBULENCE MODE IS SOLENOIDAL OR COMPRESSIVE** determines which parts of a cloud reach densities high enough to form stars. The mode also determines how efficiently the stars form. In these two simulations, stars and star clusters are shown as dark points, which form primarily in the densest parts of the clouds (yellow to white). Turbulence-regulated theory reproduces not only the star-formation rate in the two cases but the factor of 15–20 higher rate in the purely compressive simulation compared with the purely solenoidal simulation.

The nature of the IMF has far-reaching applications and consequences. Astronomers need the IMF to interpret the color, brightness, and star-formation activity of all galaxies in the universe. Because stellar feedback powers the stellar life cycle and, with it, the elemental abundances of galaxies, the IMF is a central ingredient in understanding galaxy formation and evolution. Equally important is the role of star formation in the origin of planets and life. In the swirling, dusty gas disks around young stars, the radiation, gravity, and mass of the star—and thus the IMF—control the formation of planets.

To understand the IMF, we must figure out how it depends on the characteristics of the stars' parental molecular clouds. We also need sophisticated numerical techniques that incorporate a wide range of physical processes: gravity, turbulence, magnetic fields, and stellar feedback—all at the highest numerical resolution currently available with supercomputer technology. To compare theory with observations, we need radiative-transfer calculations and other so-called forward modeling that make it possible to match physical quantities, such as a cloud's three-dimensional gas density, with observationally accessible quantities, such as a two-dimensional image of the IR emission from a particular molecular transition.

A promising theoretical framework for calculating the IMF was put forward in 2009 by Patrick Hennebelle and Gilles Chabrier.[9] It was based on the statistics of supersonic turbulence and the resulting density distribution. A key but untested prediction of their theory is that the peak of the IMF—that is, the characteristic mass of most stars—is sensitive to the nature of the turbulence. Recent numerical studies of the IMF remain inconclusive because of low number statistics, in that too few stars formed in the simulations to obtain a statistically meaningful distribution; the absence of radiative feedback; the absence of magnetic fields; or a turbulent velocity field that was fixed in its initial state rather than allowed to evolve. Moreover, nearly all the current simulations that seek to model the IMF have star-formation rates at least an order of magnitude higher than typically observed, because of the isolated boundary conditions or because of the absence of magnetic fields, stellar feedback, or both. Those shortcomings lead to gravitational collapse that proceeds too quickly.

Another critical problem arises from the accuracy and convergence of the numerical methods. The simulation resolution required to obtain physically meaningful results is extremely hard to achieve in studies of the IMF. That limitation will be obviated once our existing modeling capabilities for gravity, turbulence, and magnetic fields are supplemented with new and sophisticated subresolution techniques, so-called subgrid models, for jet and outflow feedback and for radiation feedback.

However, even with subgrid models, significant challenges persist. First and foremost, jet launching and other small-scale processes originate on the scales of the protostar, which is just a few stellar radii, yet they determine the evolution of gas cores and clouds on much larger scales, about 50 million times bigger than the star. The dynamic range is so enormous that it cannot be spanned from first principles with current modeling techniques.

For example, current subgrid models do not take into account the launching of multiple interacting jets and outflows from binary star systems. The impact of the jets and outflows, shown in figure 3, on the growth of protostars was quantified in the 2017 study by Rajika Kuruwita and colleagues.[10] The panels show the formation of a single star, a tight binary, and a wide binary. We see that the launching of jets and outflows strongly depends on how many protostars are in the disk and what their separation is. Of the three cases, the single protostar launches the most powerful jet,

**MOVIES OF THE SIMULATIONS SHOWN IN FIGURES 1, 2, AND 3 ARE AVAILABLE AT**
http://goo.gl/sRzCzK
http://goo.gl/EBeVoR
AND
http://goo.gl/jTh49d

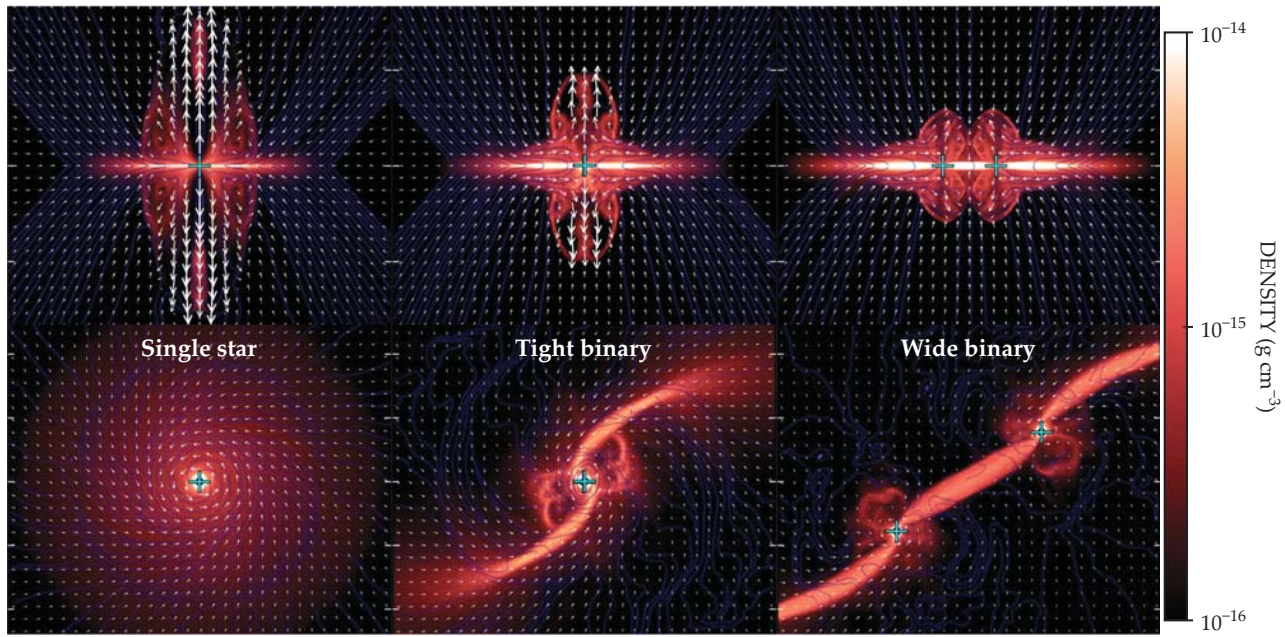

**FIGURE 3. THE JETS AND OUTFLOWS THAT STREAM FROM A PROTOSTELLAR DISK** feed material back into the cloud from which the disk and protostar formed. In these simulations, arrows indicate the velocity fields and blue crosses indicate the positions of the protostars. The jets are strongest when only one protostar is formed and weakest when two stars form in a wide binary. The simulations demonstrate that small-scale effects, such as the multiplicity of star formation, influence the amount of stellar feedback on much larger scales.

followed by the tight binary, and then the wide binary. The difference in outflow efficiency is a result of the perturbations that the tight and wide binaries induce in the disk (the central bar-like structures and in the spiral arms in the two binary cases). The perturbations disrupt the coherence of the magnetic field and reduce the disks' ability to launch magneto-centrifugal jets. Consequently, tight and wide binaries in Kuruwita and her colleagues' simulations could accrete more mass more quickly than the single star could: 10% and 20%, respectively, only 3000 years after their initial formation.

### First stars in the universe

We already know that all the scales involved in forming structure in the universe—from galaxy clusters down to planetary systems—are linked by a chain of physical processes. As yet, astronomers lack a comprehensive approach to structure formation that successfully combines the huge range of scales involved into a coherent picture. (For a discussion of previous simulation techniques used for modelling the first stars, see Tom Abel's article, "The first stars, as seen by supercomputers," PHYSICS TODAY, April 2011, page 51.)

Previous simulations indicated that the first stars may have formed with a range of masses. As a halo collapses, it produces not only a massive central star[11] but also an accretion disk that fragments into low-mass companion stars.[12] Crucially, those early studies omitted two important physical phenomena that were present in the early universe and which we know retard star formation in today's universe: magnetic fields[13] and feedback from jets and other outflows, which can reduce the average stellar mass to a third of what it would be without the feedback.[14] The significant reduction in the stellar mass by the combined effects of magnetic fields and jet-outflow feedback may have a critical impact on the mass function of the first stars and, therefore, on the ionizing radiation and chemical enrichment produced by them.

New computational techniques have been developed over the past few years, along with powerful supercomputers that combine thousands of computer cores. Such tools will ultimately enable us to determine the origin of stellar masses from the epoch when the first stars reionized the cosmos to the present day, when stars control the number and amount of heavy elements and star dust that ultimately enable the formation of planets and life.

### REFERENCES

1. M.-M. Mac Low, R. S. Klessen, *Rev. Mod. Phys.* **76**, 125 (2004); B. G. Elmegreen, J. Scalo, *Annu. Rev. Astron. Astrophys.* **42**, 211 (2004); C. F. McKee, E. C. Ostriker, *Annu. Rev. Astron. Astrophys.* **45**, 565 (2007); P. Padoan et al., in *Protostars and Planets VI*, H. Beuther et al., eds., U. Arizona Press (2014), p. 77.
2. C. Federrath, R. S. Klessen, *Astrophys. J.* **761**, 156 (2012).
3. C. Federrath et al., *Astrophys. J.* **832**, 143 (2016).
4. A. T. Barnes et al., *Mon. Not. R. Astron. Soc.* **469**, 2263 (2017).
5. S. E. Meidt et al., *Astrophys. J.* **779**, 45 (2013).
6. J. H. Orkisz et al., *Astron. Astrophys.* **599**, A99 (2017).
7. K. Jin et al., *Mon. Not. R. Astron. Soc.* **469**, 383 (2017); B. Körtgen, C. Federrath, R. Banerjee, *Mon. Not. R. Astron. Soc.* **472**, 2496 (2017).
8. S. S. R. Offner et al., in ref. 1, *Protostars and Planets VI*, p. 53.
9. P. Hennebelle, G. Chabrier, *Astrophys. J.* **702**, 1428 (2009).
10. R. L. Kuruwita, C. Federrath, M. Ireland, *Mon. Not. R. Astron. Soc.* **470**, 1626 (2017).
11. T. Abel, G. L. Bryan, M. L. Norman, *Science* **295**, 93 (2002).
12. P. C. Clark et al., *Science* **331**, 1040 (2011); T. H. Greif et al., *Astrophys. J.* **737**, 75 (2011); H. Susa, K. Hasegawa, N. Tominaga, *Astrophys. J.* **792**, 32 (2014); S. Hirano et al., *Mon. Not. R. Astron. Soc.* **448**, 568 (2015).
13. J. Schober et al., *Astrophys. J.* **754**, 99 (2012).
14. C. Federrath et al., *Astrophys. J.* **790**, 128 (2014).
15. C. Federrath, *Mon. Not. R. Astron. Soc.* **450**, 4035 (2015). PT